\documentclass[manuscript]{aastex}
\usepackage{natbib}
\usepackage{lineno}
\usepackage[figuresright]{rotating}
\usepackage{lscape}

\shorttitle{Evolving spectral-timing properties of a type-I X-ray burst of 4U~1608--52}
\shortauthors{Chen et al.}

\begin{document}

%%%%%%%%%%%%%%%%%%%%%%%%%%%%%%%Title%%%%%%%%%%%%%%%%%%%%%%%%%%%%%%%%%%%%%

\title{Insight-HXMT observation on 4U~1608--52: evidence of  interplay between   thermonuclear burst  and     accretion environments}

%\author{Y. P. Chen$^{1}$,  S. Zhang$^{1,2}$, S. N. Zhang$^{1,2}$, L. Ji$^{3}$, L. D. Kong$^{1,2}$, P. J. Wang$^{1,2}$}

\author{Yu-Peng Chen\textsuperscript{*}}
%\email{chenyp@ihep.ac.cn}
\affil{Key Laboratory for Particle Astrophysics, Institute of High Energy Physics, Chinese Academy of Sciences, 19B Yuquan Road, Beijing 100049, China}

\author{Shu Zhang\textsuperscript{*}}
%\email{szhang@ihep.ac.cn}
\affil{Key Laboratory for Particle Astrophysics, Institute of High Energy Physics, Chinese Academy of Sciences, 19B Yuquan Road, Beijing 100049, China}

\author{Long Ji}
\affil{School of Physics and Astronomy, Sun Yat-Sen University, Zhuhai, 519082, China}

\author{Shuang-Nan Zhang}
%\email{zhangsn@ihep.ac.cn}
\affil{Key Laboratory for Particle Astrophysics, Institute of High Energy Physics, Chinese Academy of Sciences, 19B Yuquan Road, Beijing 100049, China}
\affil{University of Chinese Academy of Sciences, Chinese Academy of Sciences, Beijing 100049, China}

\author{Ling-Da Kong}
%\email{kongld@ihep.ac.cn}
\affil{Key Laboratory for Particle Astrophysics, Institute of High Energy Physics, Chinese Academy of Sciences, 19B Yuquan Road, Beijing 100049, China}
\affil{University of Chinese Academy of Sciences, Chinese Academy of Sciences, Beijing 100049, China}

\author{Peng-Ju Wang}
\affil{Key Laboratory for Particle Astrophysics, Institute of High Energy Physics, Chinese Academy of Sciences, 19B Yuquan Road, Beijing 100049, China}
\affil{University of Chinese Academy of Sciences, Chinese Academy of Sciences, Beijing 100049, China}

\author{Zhi Chang}
\affil{Key Laboratory for Particle Astrophysics, Institute of High Energy Physics, Chinese Academy of Sciences, 19B Yuquan Road, Beijing 100049, China}

\author{Jing-Qiang Peng}
\affil{Key Laboratory for Particle Astrophysics, Institute of High Energy Physics, Chinese Academy of Sciences, 19B Yuquan Road, Beijing 100049, China}
\affil{University of Chinese Academy of Sciences, Chinese Academy of Sciences, Beijing 100049, China}

\author{Jin-Lu Qu}
\affil{Key Laboratory for Particle Astrophysics, Institute of High Energy Physics, Chinese Academy of Sciences, 19B Yuquan Road, Beijing 100049, China}
\affil{University of Chinese Academy of Sciences, Chinese Academy of Sciences, Beijing 100049, China}

\author{Jian Li}
\affil{CAS Key Laboratory for Research in Galaxies and Cosmology, Department of Astronomy, University of Science and Technology of China, Hefei 230026, China}
\affil{School of Astronomy and Space Science, University of Science and Technology of China, Hefei 230026, China}

%\affil{$^{1}$ Key Laboratory of Particle Astrophysics, Institute of High Energy Physics, Chinese Academy of Sciences, Beijing 100049, China}
%\affil{$^{2}$ University of Chinese Academy of Sciences, Chinese Academy of Sciences, Beijing 100049, China}

\email{chenyp@ihep.ac.cn, szhang@ihep.ac.cn}

%\altaffiltext{5}
%{
%%School of Physical Sciences, Dublin City University, Glasnevin, Dublin 9, Ireland
%}

%%%%%%%%%%%%%%%%%%%%%%%%%%%%%%% Abstract %%%%%%%%%%%%%%%%%%%%%%%%%%%%%%%%%
\begin{abstract}
%A type-I burst could have an influence upon the accretion process through radiation pressure and compton scattering  both for the accretion disk and the corona/boundary-layer of an X-ray binary, and vice versa.
A type-I burst could influence the accretion process through radiation pressure and Comptonization both for the accretion disk and the corona/boundary-layer of an X-ray binary, and vice versa.
We investigate the temporal evolution of a bright photospheric radius expansion (PRE) burst of 4U~1608--52 detected by Insight-HXMT in 1--50 keV, with the aim of studying the interplay between the burst and persistent emission. Apart from the emission from the neutron star (NS) surface,  we find the residuals both in the soft ($<$3 keV) and hard ($>$10 keV) X-ray band. %  in which the hard X-ray excess is rarely reported in the literature.
Time-resolved spectroscopy reveals that the excess can be attributed to an enhanced pre-burst/persistent emission or the Comptonization of the burst emission by the corona/boundary-layer.
The Comptonization model is a convolution thermal-Comptonization model (thcomp in XSPEC) and the Comptonization parameters are fixed at the values derived from the persistent emission.
%and no iron emission line is observed around 6 keV.
%We find the NS apparent radius at the decaying phase are a factor of 1.28 bigger than that at the rising phase.
 We  find, during the PRE phase, after the enhanced pre-burst/persistent emission or the  Comptonization of the burst emission is removed, the NS surface emission shows a plateau, and then a rise until the photosphere touches down to the NS surface, resulting in a flux peak at that moment.
We speculate that %the radius difference between the rising and decay phase and flux peak
the findings above correspond to that the obscured lower part of the NS surface by the disk is exposed to the line of sight due to the inner disk evaporation by the burst emission. %This findings are absent using the convolution thermal-Comptonization  model to fit the burst spectra.
The consistency between the $f_{a}$ model and  convolution thermal-Comptonization
 model indicates the interplay between  thermonuclear bursts and accretion environments.
These phenomena  did not usually show up by the conventional blackbody model fitting, which may be due to  low count rate and narrow energy  coverage in previous observations.

%Under this assumption, the inclination angel are  derived at ? degree.
%But other than that,  in previous literature, there are other   NS-surface-emission profiles in the PRE phase, observed by Insight-HXMT and NICER. We speculate that difference profile maybe related to the different disk/corona struction  or location during outbursts evolution.

\end{abstract}

%%%%%%%%%%%%%%%%%%%%%%%%%%%%%%%%% Section 1 %%%%%%%%%%%%%%%%%%%%%%%%%%%%%%%
\section{Introduction}
Type-I X-ray burst, also named thermonuclear bursts, are triggered by  unstable thermonuclear burning of the accreted fuel  from a  low-mass X-ray binary (LMXB)  hosting a neutron star (NS) (for reviews, see \citealp{Lewin,Cumming,Strohmayer,Galloway}).
Since its first detection in 1975 from 3A~1820--30, so far there are 116 Galactic X-ray binaries observed to produce thermonuclear bursts\footnote{https://personal.sron.nl/$\sim$jeanz/bursterlist.html},      manifesting  a sudden increase in the X-ray luminosity followed by an exponential decay and with a typical duration  about tens seconds.
The most luminous bursts are the photospheric radius expansion (PRE) events,
for which the peak flux reaches the Eddington luminosity of the NS.

Among some of the thousands of observed bursts \citep{Galloway2020},
  observations on bursters by RXTE \citep{int2013,Worpel2013,Ball2004,KeeK2014b}, INTEGRAL \citep{Sanchez2020}, NICER \citep{Keek2018,Keek2018a}, AstroSat \citep{Bhattacharyya2018} and Insight-HXMT \citep{chen2018,chen2019},
revealed interactions between the burst emission and the accretion environment:  the continuum spectrum was observed to have  an enhancement at soft X-ray and/or a shortage at hard X-rays \citep{Worpel2013,Worpel2015,chen2012,ji2013}.  Such spectral deviations are considered as burst-induced and might be relevant to disk reflection, accretion rate increase, and corona cooling \citep{Ball2004,KeeK2014b,Degenaar2018}.

Moreover, the reflection spectrum, consisting of  discrete lines and a hump peaking at 20--40 keV,  is interpreted as disk refection of an illuminant from the corona/boundary-layer.
The burst emission  could also serve as the illuminant to the disk, and a reflection component is correspondingly observed during the burst.
However, so far only iron line is firmly detected during bursts, specifically during the long-duration superbursts \citep{Ball2004,KeeK2014b}.
The  observations above are the influence of the bursts on the accretion environment; however,
there are few observational results reported related to the burst spectral change caused by the accretion disk/corona.

%Superbursts are the most energetic type-I X-ray bursts. They have typical duration of hours, significantly longer than the bursts detected from the 111 Galactic burst sources\footnote{https://personal.sron.nl/~jeanz/bursterlist.html}. Superbursts are most likely fueled by burning of carbon ashes left over by the previous stable burning helium. The superbursts from 4U~1820--30 \citep{Ball2004} and 4U~1636--536 \citep{Keek2014a} are the only two   with high data quality with respect to all others  detected so far \citep{Keek2012}. Abundant phenomena include but are not limited to,  increase of local absorption column density,   deviation from  pre-burst emission and discrete line, which were reported during the superbursts.

4U~1608--52 is a prolific burster  located at the Galactic plane \citep{Belian}.
More than 100 type-I X-ray bursts, inhabited its outbursts which have a typical frequency of once per 1--2 years since its discovery, were regularly observed.
The distance was estimated as $D\sim$2.9--4.5 kpc  based on the peak flux pf PRE bursts  $\sim$1.2--1.5  $\times10^{-7}~{\rm erg}~{\rm cm}^{2}~{\rm s}^{-1}$ (e.g. \citealp{Galloway,Poutanen}).
Its spin is around  $\nu$=619 Hz \citep{Muno,Galloway}, based on the burst oscillation detection.
% with a burst oscillation revealed at frequency around  $\nu$=619 Hz \citep{Muno,Galloway}. However, no pulsation is visible in the continuum emission during the outbursts.

In this present investigation, we provide a broad-band spectral
view of 4U~1608--52 during  its 2020 outburst observed by NICER and Insight-HXMT, both for its outburst and burst emission.
We first  describe the data reduction procedure of NICER and Insight-HXMT in Section 2. We then present an
in-depth spectral analysis and model parameters of its outburst emission  in Section 3.1,  burst lightcures and spectral evolution in Section 3.2 and Section 3.3.
 Finally, we summarize our results and discuss their implications in Section 4.

\section{Observations and Data reduction}

\subsection{Insight-HXMT}
 Hard X-ray Modulation Telescope (HXMT, also
dubbed as Insight-HXMT, \citealp{Zhang2020}) excels in its broad energy band (1--250 keV)  and a large effective area in hard X-rays energy band. \textbf{It carries three collimated telescopes: the High Energy X-ray Telescope (HE; poshwich NaI/CsI, 20–250 keV, $\sim$ 5000 cm$^2$), the Medium Energy X-ray Telescope (ME; Si pin detector, 5–40 keV,  952 cm$^2$) and the Low Energy X-ray telescope (LE; SCD detector, 1–12 keV, 384 cm$^2$).}
Under the quick read-out system of Insight-HXMT detectors, there is little pile-up effect event at the PRE burst peak. Insight-HXMT Data Analysis software (HXMTDAS) v2.04 are used to analyze the data.
 The data are reduced following
the recommended procedure of the Insight-HXMT Data Reduction
Guide v2.04\footnote{http://hxmtweb.ihep.ac.cn/SoftDoc.jhtml}, which are screened in the standard criterion include in Insight-HXMT pipelines: lepipeline, mepipeline and hepipeline.

Two bursts were detected by \textbf{three-payloads of}  Insight-HXMT from 4U~1608--52 in its 2020 outburst, as shown in Table \ref{tb}.
\textbf{However, the second burst, which is only half bright as the first one, did not fall into the good-time-interval of LE; }
%For   the second  burst, LE does not have good time interval (GTI);
thus we only analyzed the first burst occurred at MJD 59069.770768 in this work.

%As shown in Fig. \ref{fig_outburst_lc}, the contemporary count rates of MAXI and BAT are $\sim$ 75  mCrab and $\sim$ 25 mCrab respectively, suggesting that the burst is located at the soft state of the outburst.
%Since the detailed background model is still in progress,
%Because its faint persistent/outburst flux and high background of Insight/HXMT, only the spectra and lightcurves of the bursts were extracted.

%The background spectra and response files are extracted by lebkgest and lerspgen

 The persistent spectra adapted the GTI exclude the time span before the burst peak time 100 s and after the burst peak time 200 s.
The persistent spectra of LE are rebinned by ftool grppha with minimum of 100 counts per grouped bin. For the ME, the spetcra are  binned up by a factor 20, due to its background  is comparable with the source emission.

\begin{figure}[t]
\centering
   \includegraphics[angle=0, scale=0.5]{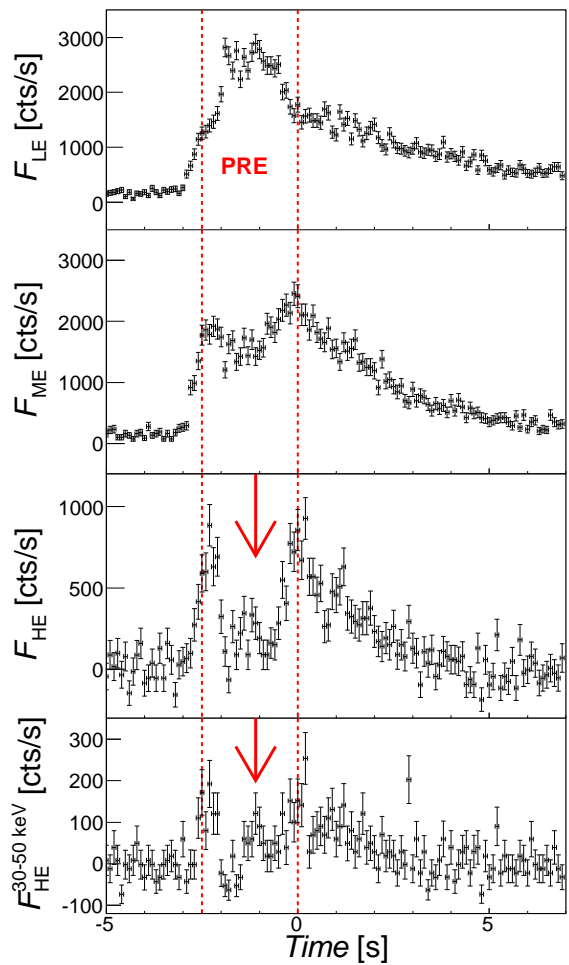}
 \caption{
 Net lightcurves of the  type-I X-ray burst detected in the Insight-HXMT observation of 4U~1608--52 with time bin 0.1 s. The top, 2nd, 3rd  and bottorm panel is LE, ME, HE results in their full energy band and HE in 30--50 keV respectively.
 The red lines indicates the PRE phase. The arrows indicates the hard X-ray excess during the PRE phase.
  % Background is subtracted.
  }
\label{fig_burst_lc}
\end{figure}

For the burst, we perform the time-resolved spectroscopy with a time resolution of 0.25 s, and define the time of the bolometric flux peak as a time reference (0 point in Fig. \ref{fig_burst_lc} and Fig. \ref{fig_fit}).
The burst spectra are rebinned by ftool grppha with minimum of 10 counts per grouped bin.

 In the calibration experiments on ground and the first two years in orbit, the recommend energy band for spectral fitting of LE is 1-10 keV, except very bright sources with a flux brighter than several Crab. After the midyear of 2019, the recommend band is shrunken to 2-10 keV, which mostly dues to the poor background estimation because of an increase of detector temperatures and something else. However, in the burst spectral analysis, we take the pre-burst emission as background, so we extend the energy band to 1-10 keV in burst spectra fitting, but still adapt 2-10 keV in persistent emission spectral fitting.
 \textbf{
 We notice that, the LE spectral residuals  of the persistent emission have rather complex structures in the energy band $<$ 1.5 keV, where the bursts have only a few data points.  % For comparison,  we also display the LE spectrum of the persistent emission in the 1-2 keV energy range.  The shape of the persistent residuals  conduct a steep, precipitous cliff, which is different with the burst results. Moreover, the persistent residuals mostly locate $<$ 1.5 keV; but there are a few data points in this band for the burst emission.
 }
 We also use the 2-10 keV to analyze the burst spectra, and get the roughly consistent results within parameter’s error bar.
It is  a similar consideration that we extend the ME band to 8--30 keV and the HE band to 25--50 keV in the burst spectra fitting.
In short, for the persistent emission spectra fitting of LE and ME, the energy bands are limited to    2--10 keV and 10--20 keV; for the burst  spectra fitting of LE, ME and HE, the energy bands used are 1--10 keV and 8--30 keV and 25--50 keV respectively.

\begin{figure}[t]
\centering
      \includegraphics[angle=0, scale=0.3]{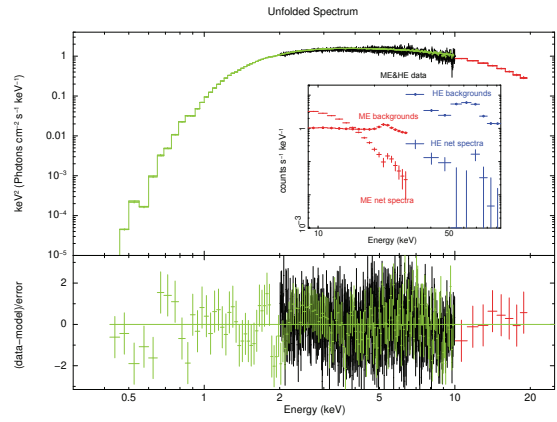}
 \caption{The spectral fit results of the  persistent emission by  LE (black), ME (red) and NICER (green)  with model cons*tbabs*thcomp*diskbb.
 \textbf{The embedded panel shows the background level for ME (red) and HE (blue).}
 }
\label{sep_nicer}
\end{figure}

During fittings of the persistent emission, the LE data in 2--10 keV and ME data in 10--20 keV are used,  while the ME data $>$ 20 keV and the HE spectra are not used to fit  because of faint source flux and strong background.  \textbf{As shown in Fig. \ref{sep_nicer},  the persistent emission detected by ME in 20--30 keV and HE in 25--100 keV is very weak compared with the background. Most of these spectral channels are close to or fainter than the systematic uncertainty of the background (1\%); e.g., for the count rate detected by HE in 30--50 keV, the background of $\sim$120 cts/s is comparable to the detected count rate of $\sim$123 cts/s. Thus the persistent emission has a count rate of $\sim$ 3 cts/s which is close to the systematic uncertainty of the estimated background. }
In addition, we added a systematic uncertainty of 1\% to the Insight-HXMT spectra in 1--100 keV, to account for   systematic uncertainties in the detector calibrations  \citep{Li2020}.

%The background spectra and response files are extracted by lebkgest and lerspgen

\subsection{NICER}
  On 2020 August 8, within the same day when Insight-HXMT detected the burst from 4U 1608--52, NICER also observed the same source.  The OBSID is 3657026501, with a good time interval $\sim$ 2 ks and a count rate $\sim$ 900 cts/s in the 0.3–12 keV band.   %as shown in Figure  \ref{fig_lc_nicer_le_me}.
  However, the NICER missed the type-I X-ray burst because of an observation gap.

 The NICER data are reduced using  the  pipeline tool  nicerl2\footnote{https://heasarc.gsfc.nasa.gov/docs/nicer/nicer\_analysis.html} in NICERDAS v7a  with the standard NICER filtering  and using  ftool  XSELECT to extract lightcurves and spcectra. The background is estimated using the tool nibackgen3C50 \citep{Remillard2022}. The Focal Plane Module (FPM) No. 14 and 34 are removed from the analysis  because of increased detector noise. The response matrix files (RMFs) and   ancillary response files (ARFs) are generated   with the ftool nicerrmf and nicerarf.  The spectra are rebinned by ftool ftgrouppha \citep{Kaastra2016} optimal binning algorithm and plus minimum of 25 counts per grouped bin. Other rebin method, e.g., minimum of 100 counts per grouped bin by ftool grppha, are adapted. As expected, the fit results are consistent with each other within parameter’s error bar.

  For the ISM absorption, we use tbabs in the spectral model and   wilm abundances \citep{Wilms2000}. To erase the residuals $<$ 1 keV, three absorption edges are added in spectra fitting: 0.56 keV, 0.71 keV, and 0.87 keV.   We added a systematic uncertainty of 1\% to the NICER spectra.

 %As shown in Fig. \ref{fig_lc_nicer_le_me},
From NICER and Insight-HXMT lightcurves, the none-burst/persistent emission is stable in our observations.
We jointly fit the persistent spectra observed with NICER and Insight-HMXT, as shown in Fig. \ref{sep_nicer}.
The joint fit of the spectra covers an energy band of 0.4--10 keV, 2--10 keV and 10--20 keV for NICER, LE and ME, respectively.

The spectra are fitted with XSPEC v12.11.1 and the model parameters are estimated with a 68\% confidence level (1 $\sigma$).

\section{Analysis and Results}
%0.63 cts/s 13.6-12.968599
\subsection{None-burst/persistent emission detected by NICER and Insight-HXMT}

 The jointed NICER and Insight-HXMT data in a broader energy range 0.4--20 keV, give us an opportunity  to utilize a more physics meaningful model to fit the persistent emission, rather than the simplified models, i.e., a simple photon power law, a power law  with high energy exponential rolloff (cutoffpl in xspec) and a broken power law (bknpow in xspec).
We fit the joint of NICER and Insight-HXMT (LE and ME) spectrum with an  absorbed convolution
  thermal Comptonization model (with an input seed photon spectrum  diskbb), available as thcomp (a more accurate version of nthcomp) \citep{Zdziarski2021} in XSPEC,
 which is described by the optical depth $\tau$, electron temperature $kT_{\rm e}$, scattered/covering fraction $f_{\rm sc}$.

 % and assumes a spherical distribution of the thermal electrons for the Comptonization.

The  hydrogen column (tbabs in XSPEC) %is fixed at 1.5$\times$10$^{22}~{\rm cm}^{-2}$  \citep{Penninx1989},
 accounts for both the line of sight column density and as well any intrinsic absorption near the source.
 %and corresponds to the best-fit value for the persistent spectrum.
 The seed photons are in a shape of diskbb, since the thcomp model is a convolution model and the fraction of Comptonization photons is  also given in the model.
 %from any form of the seed spectrum.
 %is provided as a convolution function, and thus it can be used with any form of seed photons.  This is a convolution model, allowing for Comptonization of a fraction of photons from any form of the seed spectrum. We assume the seed photons have a disk blackbody spectrum (diskbb; Mitsuda et al. 1984).

%In addition, a diskbb component  .
Normalization constants are included during fittings to take into account the inter-calibrations of the instruments.
 We keep the normalization factor of the LE data with respect to the ME and NICER data to unity.
%During the spectral analysis, a 1\% systematic error is added to account for the uncertainties of the background model and  calibration.

 Using the model above, we find an acceptable fit:$\chi_{\upsilon}$=0.95 (d.o.f. 846; Fig. \ref{sep_nicer} and Table  \ref{persist_fit}), with  the inner disc radius $R_{\rm diskbb}$ and scattered/covering fraction $f_{\rm sc}$ are found to be $\sim 15.1_{-0.8}^{+0.9}$ km (with distance 4 kpc and inclination angel 40 degree) and $0.77_{-0.05}^{+0.05}$,
respectively.
  The  derived hydrogen column density  $N_{\rm H}$  is $\sim$1.3$\times 10^{22}~{\rm cm}^{-2}$, which agrees with values previously reported  in a range of 0.9--1.5 $\times 10^{22}~{\rm cm}^{-2}$ \citep{PenninxW1989,ArmasPadilla2017}.
The thcomp parameters,   $\tau$ and $kT_{\rm e}$ are well consistent with a previous outburst in soft state of 4U 1608--52 \citep{ArmasPadilla2017}, which derived the parameters were derived with the nthcomp model.
The inferred bolometric flux in 1--100 keV is $7.33_{-0.03}^{+0.06}\times10^{-9}~{\rm erg}~{\rm cm}^{-2}~{\rm s}^{-1}$. The constant of ME and NICER is 0.93$\pm$0.02 and 1.05$\pm$0.01  respectively.

\textbf{Using model of cons*tbabs*(diskbb+nthcomp) to fit the spectra, i.e., non-convolutional Comptonization model,  we get a similar results for both  the corona and the disk temperatures but with a smaller normalization of the disk. The shortage of the disk normalization than the convolution model is corresponding to the
missing part of the disk emission which is supposed to be scattered in the corona. }

 Another assumption that the seed photons of the Comptonization are from the NS surface, i.e., the diskbb component is substituted by a blackbody component in the aforementioned convolution model, is also attempted.
Taking this approach, spectral fits yield a roughly same thcomp parameters but with $\chi_{\upsilon}$=1.12  (the same d.o.f.)  and soft residuals $<$ 2 keV.
Furthermore, the derived blackbody radius is  $34.1\pm1.0$ km, which is far greater than the NS radius.
A hybrid model \citep{ArmasPadilla2017}, i.e., three-component model (diskbb+thcomp*bb or bb+thcomp*diskbb) is not attempted, since the above two-component model is able to fit the data.

Since there are no iron emission line or reflection bump  above 10 keV, no reflection model are used for the spectrum fitting.
%We explore several different fits, e.g., absorbed diskbb+powerlaw or diskbb+nthcompt,  and find that these models can also give  acceptable fits to the joint NICER and Insight-HXMT data.
%Other models were also attemped to fit the spectra,
%The fit results

%\subsection{Data analysis}

\subsection{Burst lightcurves by Insight-HXMT}

We show the LE/ME/HE lightcurves in Fig. \ref{fig_burst_lc} with a time resolution of 0.1 s.
The burst profiles  exhibit a  typically fast rise and slow (exponential) decay in the soft X-ray band,
and manifest a plateau in soft X-ray band (LE) and two  peaks in hard X-ray band (ME\&HE), which is a typical characteristic of a PRE burst.
%and  corresponding to the burst emission with high blackbody temperatures $\sim$ 3 keV.

 %because of low blackbody temperatures $<$ 2 keV.
 In the middle of the PRE phase with a constant luminosity $L_{\rm Edd}$,  the burst emission has the lowest blackbody temperature, which could cause a dip in the HE lightcurves.
However, interestingly, there is a peak/excess  in the HE lightcurves.
For the highest 6 points in HE lightcurves in its whole energy band (20--250 keV), the hard excess is 222.3$\pm$39.3 cts/s with 5.6 $\sigma$ detection; meanwhile, the burst emission for HE (for a blackbody with a temperature of 2.0 keV and a bolometric flux of an  Eddington luminosity) should be less than 35 cts/s in this energy band. The hard excess in 30--50 keV is 71.5$\pm$18.0 cts/s with 4 $\sigma$ detection; meanwhile, the burst emission  should be negligible with $<$ 0.1 cts/s in this energy band.

This hard X-ray excess in the lightcurve suggests that there is another provenance except for the burst, which is also visible during the burst spectra analysis below.
%the burst emission may be Comptonized like the persistent emission,

%although we are not able to verify that spectrally due to insufficient statistics.

\subsection{Broad-band spectra of burst emission by Insight-HXMT}

When we fit the burst spectra, we estimate the background using the emission before the burst, i.e., assuming the persistent emission is unchanged during the burst. %the none-burst/persistent emission is stable in our observations.
%We perform the time-resolved spectroscopy with a time resolution of 0.25 s, and  define the time of the bolometric  flux peak as a time reference (zero point in Fig. \ref{fig_fit}).
%A fixed absorption (tbabs in XSPEC), as used in pre-burst fitting, is used to fit the burst spectra.
To account for the effective area calibration deviation, a constant is added to the model.
At the first attempt, for LE, the constant is fixed to 1, the others are variable during spectra fitting. The fits indicate that most of the constants of HE  and some of the constant of ME are not convergent, owing to the low-significance data. Under this situation, the constant of ME\&HE is fixed at 1 for the combined-spectra fitting.

We follow the classical approach to X-ray burst spectroscopy by subtracting the persistent spectrum
and fitting the net spectrum with an absorbed blackbody.
In the decay phase, such a spectral model generally results in an acceptable goodness-of-fit, with a mean reduced $\chi^{2}_{\upsilon}~\sim$ 1.0 (d.o.f. 20--60).
However, we note that a significant residuals are shown below 3 keV and above 10 keV, as shown in the left panel of Fig. \ref{residual}, especially the  spectra in the PRE phase, the  reduced $\chi^{2}_{\upsilon}$  are above 1.5 (d.o.f. 60--80).
%Also the burst spectral fitting results are consistent with having a detection of a PRE from the lightcurve of Insight-HXMT.

To erase the residuals, we first consider the $f_{a}$ model. Following \citet{Worpel2013} we then include an addition component for fitting the variable persistent emission.
We assume that during the burst the spectral shape of the persistent emission is unchanged, and only its normalization (known as a $f_{a}$ factor) is changeable.
As reported earlier by RXTE and NICER,
the $f_{a}$ model provides a better fit than the conventional one (absorbed blackbody).
We compare the above two models  using the F-test. % As shown in Table \ref{table},
In some cases, the $f_{a}$  model   significantly improves the fits with a p-value $\sim10^{-5}$.

\begin{figure}[t]
\centering
      \includegraphics[angle=0, scale=0.5]{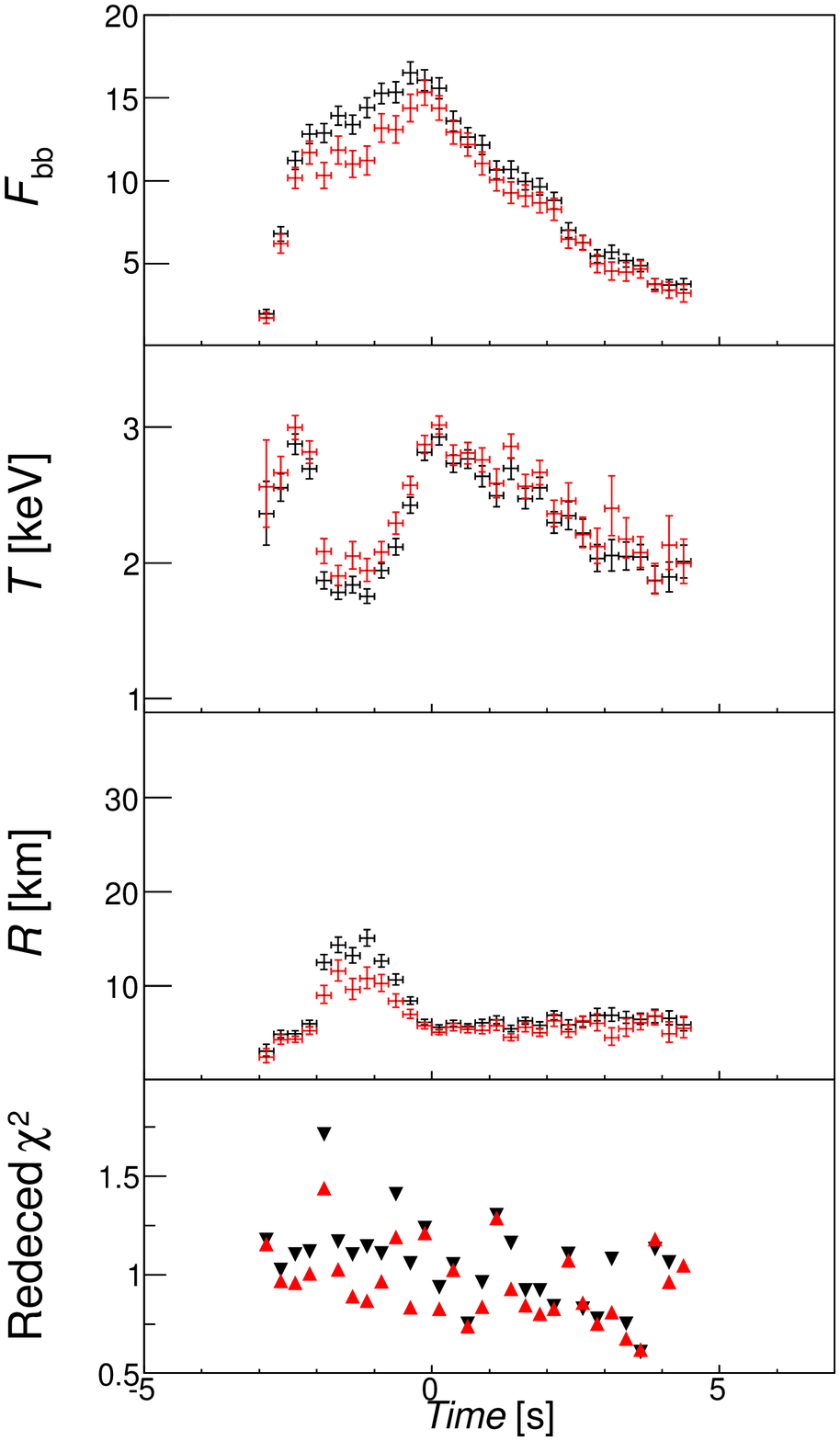}
            \includegraphics[angle=0, scale=0.5]{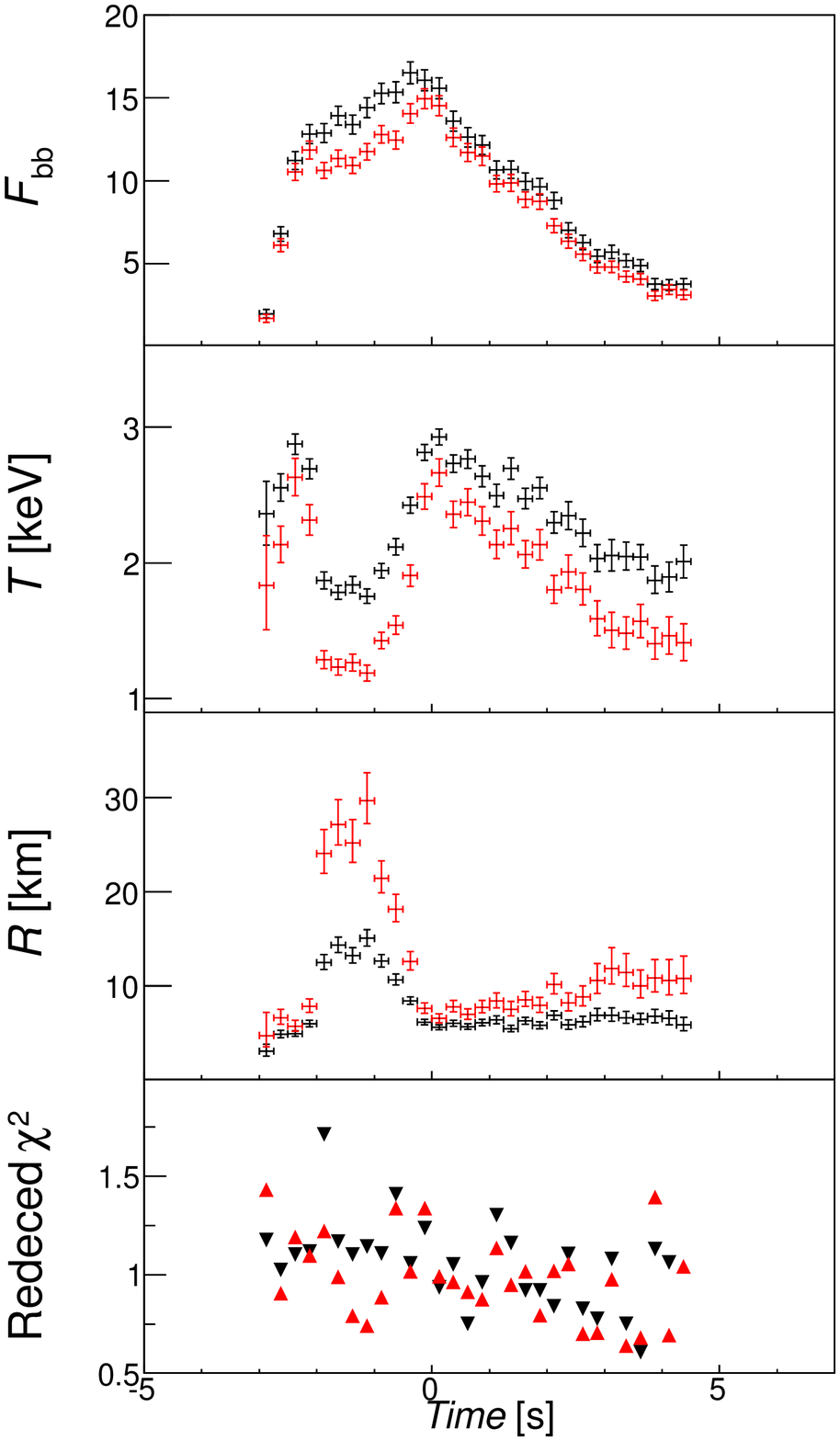}
 \caption{
Spectral fitting result of the burst with time bin 0.25 second  with  a pure blackbody (black), $f_{a}$ model (the left panel, red) and and convolution thermal-Comptonization model (the right panle, red), %pre-burst spectrum subtracted,
include the time evolution of the blackbody bolometric flux $F_{\rm bb}$, the temperature $kT_{\rm bb}$, the observed radius $R$ of NS surface at 4 kpc, the goodness of fit $\chi_{v}^{2}$.
%The black and red indicates the fitting results by a absorbed  blackbody and $f_{a}$ model, respectively.
The bolometric flux of the blackbody model $F_{\rm bb}$ is in unit of $10^{-8}~{\rm erg/cm}^{2}/{\rm s}$.  }
\label{fig_fit}
\end{figure}

\begin{figure}[t]
\centering
   \includegraphics[angle=0, scale=0.30]{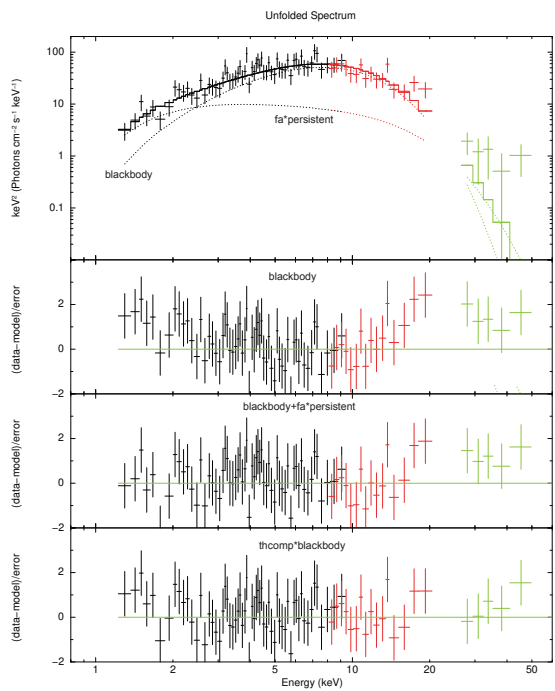}
 \caption{
 \textbf{Top panel: the spectral fits results by  LE (black), ME (red) and HE (green) when the burst reaches the maximum emission area by $f_{a}$ model, the blackbody model and enhancement of the persistent emission are labeled.
 The three panels below: residuals of spectral fits results by
 an absorbed black-body  model (the 2nd panel), $f_{a}$ model (the 3rd panel) and convolution thermal-Comptonization model (the bottom panel). }%The lower dotted curves represent  the model components for the pre-burst emission (tbabs*thcomp*diskbb).
 }
\label{residual}
\end{figure}

As shown in left panel of Figure \ref{fig_fit}, the spectral fitting results from these two models have differences mainly around the PRE phase.  By considering an additional factor $f_{a}$, the burst blackbody flux tends to slightly decrease, and the temperature becomes higher but the radius shrinks.
The $f_{a}$ factor reaches a maximum of $6.5_{-1.3}^{+1.3}$ when the radius reaches its peak.  %which indicates that the enhancement of the pre-burst emission is up to  $\sim$ 40\%$L_{Edd}$.    $L_{\rm Edd}$ is adopted from the flux at touch-down time (at the end of PRE phase when the radius has declined to the asymptotic value in the burst tail) during the burst.
During the PRE phase, the radius is up to $10.8_{-1.0}^{+1.2}$ km, which is two times larger than the radius measured at touch-down time  $5.1_{-0.3}^{+0.3}$ km (assuming a distance of 4 kpc).
This is typical  to a moderate photospheric expansion with a bolometric burst peak flux $F_{\rm bb}$  $15.3_{-0.8}^{+0.8}\times10^{-8}~{\rm erg}~{\rm cm}^{-2}~{\rm s}^{-1}$ in 0.1--100 keV.
%The bolometric burst fluence is $f_{burst}$ = 8.97 $\pm$0.08 $\times$ 10$^{-7}$ erg cm$^{-2}$ over 40 s.

Since the burst photons could also be affected by the corona/boundary-layer, we thus check if the model used in the persistent emission could be same with the burst emission.
By taking the pre-burst emission as background emission, the burst spectra are fitted by the model tbabs*thcomp*bb, in which the thcomp parameters are fixed at the persistent emission fit results.  Thus convolution  thermal Comptonization model (with an input seed photon spectra blackbody)  has the same d.o.f with the canonical blackbody model, and a more d.o.f. than the $f_{a}$ model.  The bb and thcomp represents the burst emission from the NS photosphere and  a corona/boundary-layer influence on the burst emission.
This model allows us to evaluate  the contribution from both the up-scattered by the corona/boundayr-layer and direct  photons from the NS surface.

In the PPE phase, this model provides the best fit and
yields physically acceptable spectral parameters; the obtained best-fit parameters are given in the right panel of Fig. \ref{fig_fit}. We find that this convolved  thermal-Comptonization model provides an equally good results with $f_{a}$ model but with a  more d.o.f.,  and  statistically preferred to the $f_{a}$ model   in the middle of PRE phase (with a coolest blackbody temperature). However, in the rising and decay part, such model has a bigger reduced $\chi^{2}_{\upsilon}$ than the  $f_{a}$ model and even the canonical blackbody model, which may indicates that the burst emission suffers few Comptonization during this phase.
%the corona/boundary-layer is far away from the NS surface.}

As mentioned above, the free/unfixed parameters include the blackbody temperature  $kT_{\rm bb}$ and the normalization $N_{\rm bb}$. The trend of the parameters are similar with the $f_{a}$ model, but with a greater change. Compared to the $f_{a}$ model results, the maximum radius $R_{\rm bb}$ is up to $29.5_{-2.4}^{+2.9}$ km, the minimum temperature  $kT_{\rm bb}$ is low to $1.19_{-0.06}^{+0.06}$ keV.

Other scenarios, i.e.,  burst reflection by the disk, NS atmosphere  model carbatm/hatm \citep{Suleimanov2011,Suleimanov2012,Suleimanov2018} in Xspec, are also tried to fit the burst spectra, as we did in \citet{chen2019}.
However, neither could  alleviate the residuals at soft X-ray and hard X-ray bands simultaneously.

For the hard X-ray excess detected in the lightcurve during the PRE phase, we calculate and find that the   persistent emission has not enough flux to build the enhancement.
We fake the HE spectra using the aforementioned model parameters of the persistent emission, %from jointed NICER and Insight-HXMT spectra,
the HE flux of the persistent emission model in 20--250 keV and 30--50 keV is 3 cts/s and 0.7 cts/s.
Taking into the factor  $f_{a}$ account, this model predicted enhancement flux  only equivalent to one tens of the hard excess.
The spectra residuals in hard X-ray band are also visible in the middle panel of Fig. \ref{residual} ($f_{a}$ model to fit the burst spectra in the PRE phase).
The hard X-ray excess also disfavours the reflection model because of the faint  persistent emission in hard X-ray band.

%The soft access may  be caused by the reflection  from the accretion-disk illuminated by the blackbody emission from the NS surface/photosphere during the burst. We employ  rdblur*bbrefl\_2xsolar\_0-5r.fits or relconv*bbrefl\_2xsolar\_0-5r.fits to replace the $f_{a}$ model, in which bbrefl\_2xsolar\_0-5r.fits and rdblur/relconv is corresponding to the reflection emission and   relativistic doppler broadening of the reflection component, respectively.

%In such an attempt, the fit parameters are fixed at the dimensionless spin $\alpha$ = 0.29, the disk inclination i = 38.8 degree, the inner and outer disc radii $R_{\rm in}$=6  and $R_{\rm out}$=400 in unit of $R_{\rm ISCO}$,  which are adapted from  \citet{Degenaar2015}. Since there are no hints of the iron emission line  or the reflection-bump in 20--40 keV, the ionization parameter  is pegged at the tabular boundary of log$\xi$ = 3.75. However, the additional reflection component can not alleviate the residuals at soft X-rays.

\section{Discussion}
In this work, we have presented a spectral analysis of a  PRE burst and persistent emission from 4U~1608--52 in its 2020 outburst observed by NICER and Insight-HXMT.
The persistent emission is well fitted by an absorbed convolution thermal-Comptonization model, in which 77\% disk emission is up-scattered by the corona/boundary-layer.
The X-ray burst shows  a significant spectral deviation/excess both at $<$ 3 keV and $>$ 10 keV from an absorpted blackbody in the PRE phase. %, in which the hard X-ray excess is firstly reported in the burst with short duration since 1976.
%This excess is consistent with the assumption that it is enhanced accretion rate during the burst.
%Another scenario is
This excess is consistent with that the burst emission is up-scattered by the corona/boundary-layer,  only part of the burst emission without Comptonization is detected, which mimics the Comptonization of the disk emission in the persistent emission.

\subsection{X-ray continuum}
\textbf{Based on LE\&ME lightcurves and spectral fitting results, the burst locates at the high/soft state (banana state).}
Previous works have attempt to fit the spectra with thermal (diskbb or/and blackbody) plus a Comptonization model, rather than a
convolution thermal-Comptonization model,  which will cause an underestimation of the thermal emission.
In this work, adapting thermal-Comptonization model thcomp in XSPEC, the fit results indicate that  most of the disk emission is involved in Compton upscattering.

Broadly speaking, in an accreting low-magnetic field NS, except for the emission from the NS surface, there are  at least two geometrically distinct regions to generate X-ray emission (see the review by \citealp{Done2007}): in the accretion disc  and  the boundary/spreading layer (BL/SL)  (similar to the corona of the case of an accreting black hole).
BL is supposed to spread in large radial extent in the disc midplane; whereas  SL has a  narrower spread but  spreads over a considerable height from the equatorial plane toward higher stellar latitudes.
There are claimed  judgment criteria for BL and SL based on temporal \citep{Gilfanov2003} and spectral \citep{Grebenev2002,Suleimanov2006} characteristics.

In the burst review, during the decay part of the burst, the burst emission is well fitted by a blackbody and no strong comptonization/up-scattered emission detected.
Thus the hot electrons plasma should not have a significant coverage for the NS surface.
Add that into the consideration with a big scattering factor $f_{\rm sc}$ 0.77$_{-0.05}^{+0.05}$ (the hot electrons plasma on the disk in the persistent emission), the corona-like geometry of the BL is favoured.

We find that the persistent emission is 4.8\% $L_{\rm Edd}$ and the corona/boundary-layer temperature is 3.02$_{-0.08}^{+0.08}$ keV,  which is  in the range of Comptonizing temperature expected for NS LMXBs  in the  soft state \citep{ArmasPadilla2017}.
Meanwhile, the scattering factor $f_{\rm sc}$ is  0.77$_{-0.05}^{+0.05}$,  which  is too large for the corona/boundary-layer with a lamp-post geometry. Given those above, we prefer the corona/boundary-layer with a slab/sandwich geometry, as shown in Figure \ref{fig_illustraction}.
Regarding that the  temperature and optical depth deviates from the corona canonical value,  we also prefer anther corona pattern--a so-called warm layer \citep{Zhang2000} with temperature $\sim$2--3 keV and optical depth $\sim$5--10, which is produced by  the magnetic reconnection.
The outburst spectral evolution and its understandings will be given in our forthcoming paper.

\subsection{Enhanced Persistent Emission up to 50 keV}

In several bursters,  during   the low/hard state, decrease/deficit in the hard X-ray band (30--50 keV) have been observed in  short-duration bursts which happened in the low-hard state (\citealp{chen2012} and reference therein). It is expected that the burst emission (2--3 keV), which is relatively cooler than the corona (tens keV), cause the change in the corona structure or temperature.

In this work, conversely, an enhancement hard X-ray emission are observed during a short-duration burst, which is first reported in GS~1826--238 by BeppoSAX in 30--60 keV \citep{int1999}.
\textbf{However, these two sets of  bursts located different spectral state of LMXBs.
In both cases, the soft X-ray showers of the burst may manifest as an enhancement of the input seed photons but not a sufficient cooler upon the corona.
%Compared with the bursts of GS 1826-238 by BeppoSAX, the burst of this work is brighter, i.e., the burst has more seed photons for the Comptonization by the corona/boundary. Although the corona in the soft state is cooler than that in the hard state, the HE could detect this excess by a larger effective area than BeppoSAX.
}

\textbf{
Compared with the disk component of the persistent emission, the count rate of the burst at the PRE phase is 4 times more. The emergent photons of the burst could be up-scattered to higher energy by the corona/boundary-layer.
For the comptonization of the burst during PRE phase, i.e., a  blackbody with  temperature  1.22 keV and  normalization  1.33, and a hot corona with  temperature of 3 keV, optical depth  10.2 and cover-factor  0.76, we fake a spectrum induced by the inverse Compton scattering of the blackbody emission and get a
 count rate   289 cts/s in the energy band of 30--50 keV.  Thus, the up-scattered  photons of the burst do cause  an enhancement hard X-ray emission.}

%In fact, the average 30–60 keV photon count rate of the burst in the first 11 s after the burst peak is of order halftimes that in 12–30 keV. This suggests that the burst emission may be Comptonized like the persistent emission, although we are not able to verify that spectrally due to insufficient statistics.
Based on the burst spectra fit results,
 the enhancement hard X-ray emission could be related to
%it could be related the enhancement of persistent emission or
the up-scattered burst emission by the corona/boundary-layer, just like the situation in the persistent emission, rather than enhancement accretion rate manifesting itself by elevating persistent emission with unaltered spectral shape \citep{Worpel2013,Worpel2015}.

%The low electron temperature in the spectral model of the persistent emission and faint hard X-ray emission indicates   that the burst of this work inhabits in the  soft state.
%A possible scenario is that the different accretion environment has difference influence on the interaction between the burst and the corona.

\subsection{Dynamical evolution of the disk geometry}

 As a common sense, the burst emission has an increasing   and  decreasing area during its rise and decay phase, which corresponding the hot spot spreading in the NS surface.
Meanwhile, there are at least two moments when the hot spot covers that whole NS surface, the photoshphere lift-up point and the touch-down point for the PRE burst.
\textbf{As shown in Figure 1 of \citet{Shaposhnikov2003}, the hot-spot spreads on the NS surface and then lifts up the photonsphere, i.e., from 'a' stage to 'b' stage in the Figure. There should be a moment that hot spot covers that whole NS surface in the rise phase and vice versa in the decay phase. However, there are some PRE bursts with a short increase time, i.e., that the increase time is too short for telescope to accumulating enough counts in the first moment when the hot spot covers that whole NS surface.} %Indeed, there are some PRE bursts with a sharp increase time, i.e., that the increase time is too short for telescope to detecting, e.g., the PRE bursts from 4U 1820-30.

\begin{figure}[t]
\centering
      \includegraphics[angle=0, scale=0.4]{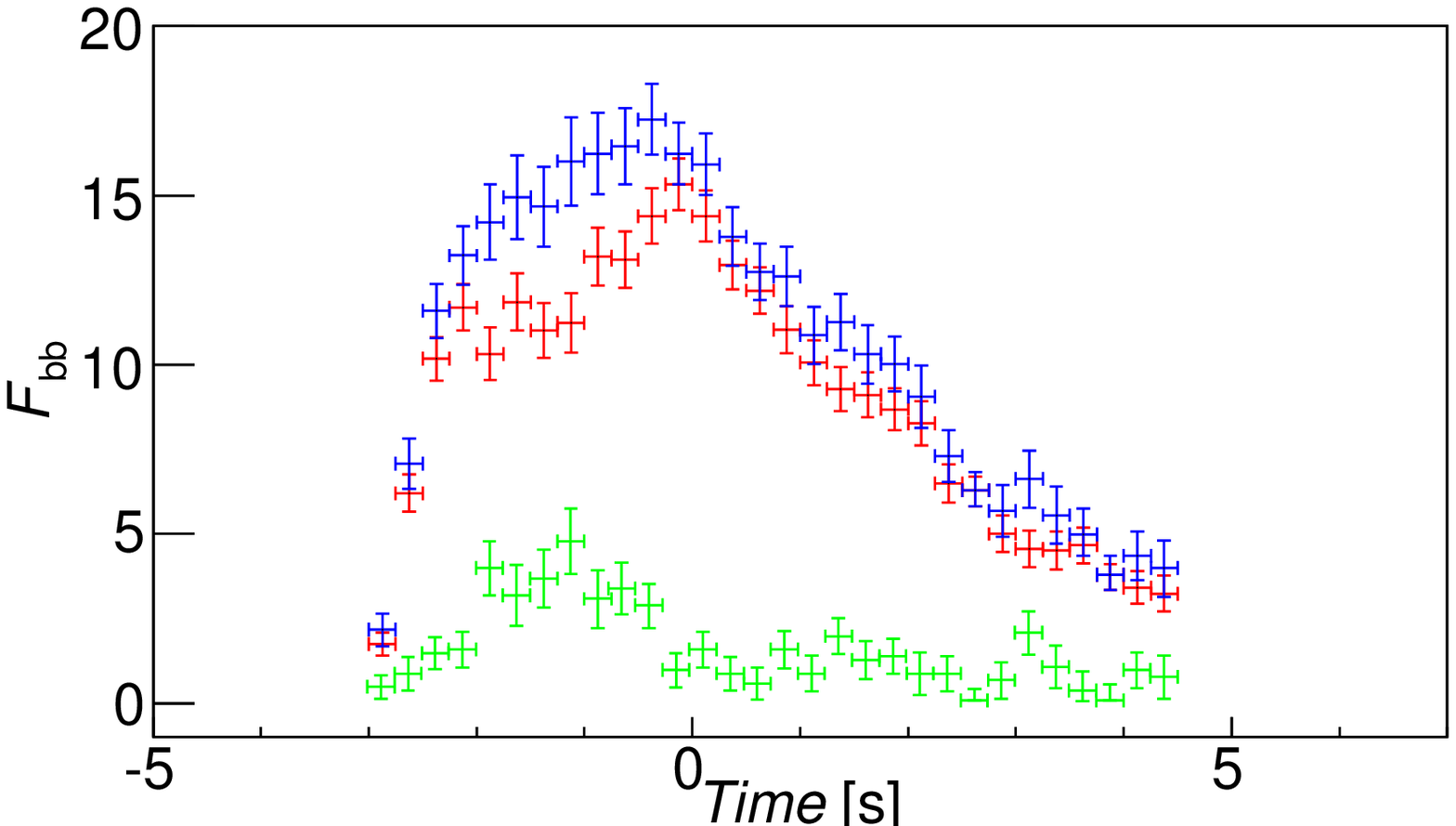}
            \includegraphics[angle=0, scale=0.4]{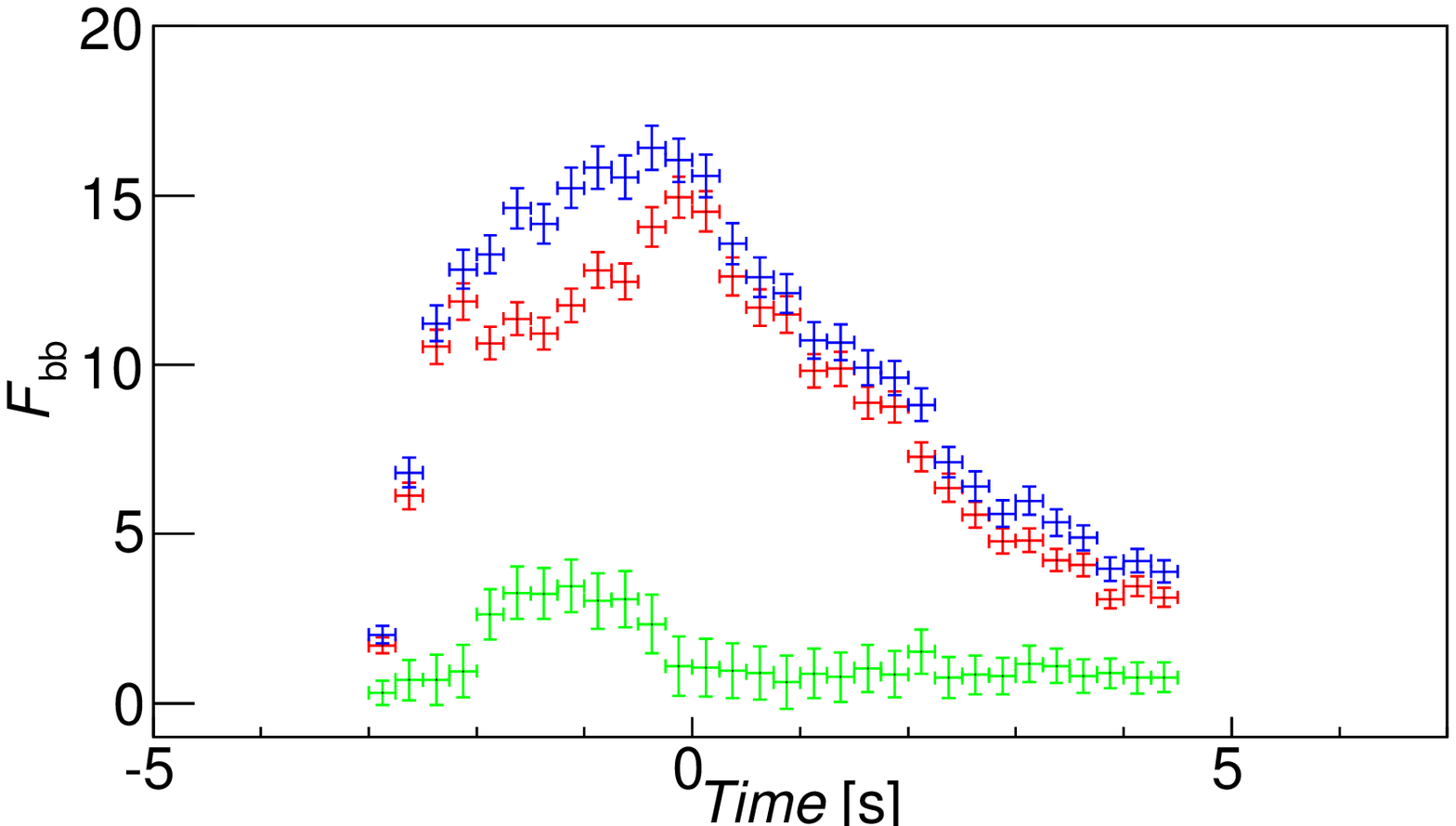}
 \caption{Bolometric flux evolution of the bursts detected by Insight-HXMT with  the $f_{a}$ model (left) and convolution thermal-Comptonization model (right). The bolometric flux (in unit of $10^{-8}~{\rm erg/cm}^{2}/{\rm s}$) evolution  of  the enhanced persistent emission (green), the burst emission (red) and the sum of the former two components (blue). %The horizontal solid lines indicate the burst flux when the photosphere are lifted from the NS surface in the PRE rising phase, which is used to compare with the burst flux in the touch-down time. %The horizontal solid line and dotted lines indicate the $L_{\rm Edd}$ and its upper/lower bound (1 $\sigma$), respectively.
  } % which is estimated by the blackbody flux of the  touch-down time during the burst.}
\label{fit_ledd}
\end{figure}

In practice, the latter is usually used to derive the NS radius. % and the former is hard to  duo to it is hard to
As shown in Figure \ref{fit_ledd}, at the touch-down point, the burst emission reaches its peak flux, both for the $f_{a}$ model and convolution thermal-Comptonization model.
A dynamical evolution of the disk geometry could cause this phenomenon, i.e., the lower NS hemisphere, which is obscured before the burst (the burst PRE phase), appears from the disk after the burst-disk interaction, as shown in Figure 1 of \citet{Shaposhnikov2003} and Figure \ref{fig_illustraction} in this work.

In theory, Poynting-Robertson drag could drain the inner-accretion-disk  by   taking away  the momentum of the accretion matter hence enlarging the local accretion rate \citep{int2013,Worpel2013,Worpel2015}, which is faster than it is being refilled \citep{Stahl,Fragile2020}.
At this moment, the inner part of the disk is hollowed out by the burst emission. % through evaporating or radiation pressure.

\begin{figure}[t]
\centering
      \includegraphics[angle=0, scale=0.4]{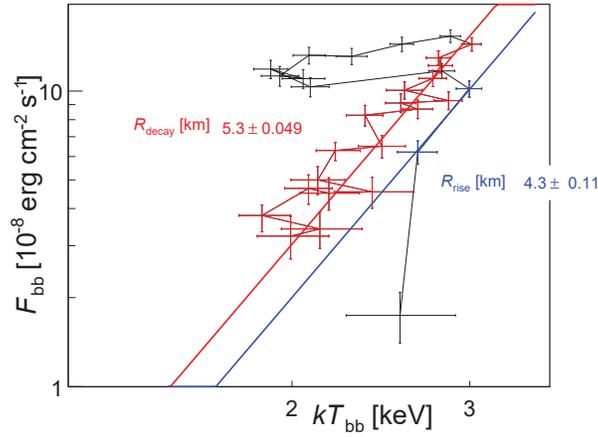}
 \caption{The flux-temperature diagram for the burst spectra in the rising (blue) and decaying (red) phase from 4U~1608--52. The diagonal blue and red lines correspond to constant radius in the rising and decaying phase. %The arrows indicate
  The time sequence is that the burst starts at the  lower right, stays at PRE phase (black) at the upper left and decays at the lower left. }
\label{fig_t_f}
\end{figure}

\begin{figure}[t]
\centering
      \includegraphics[angle=0, scale=0.22]{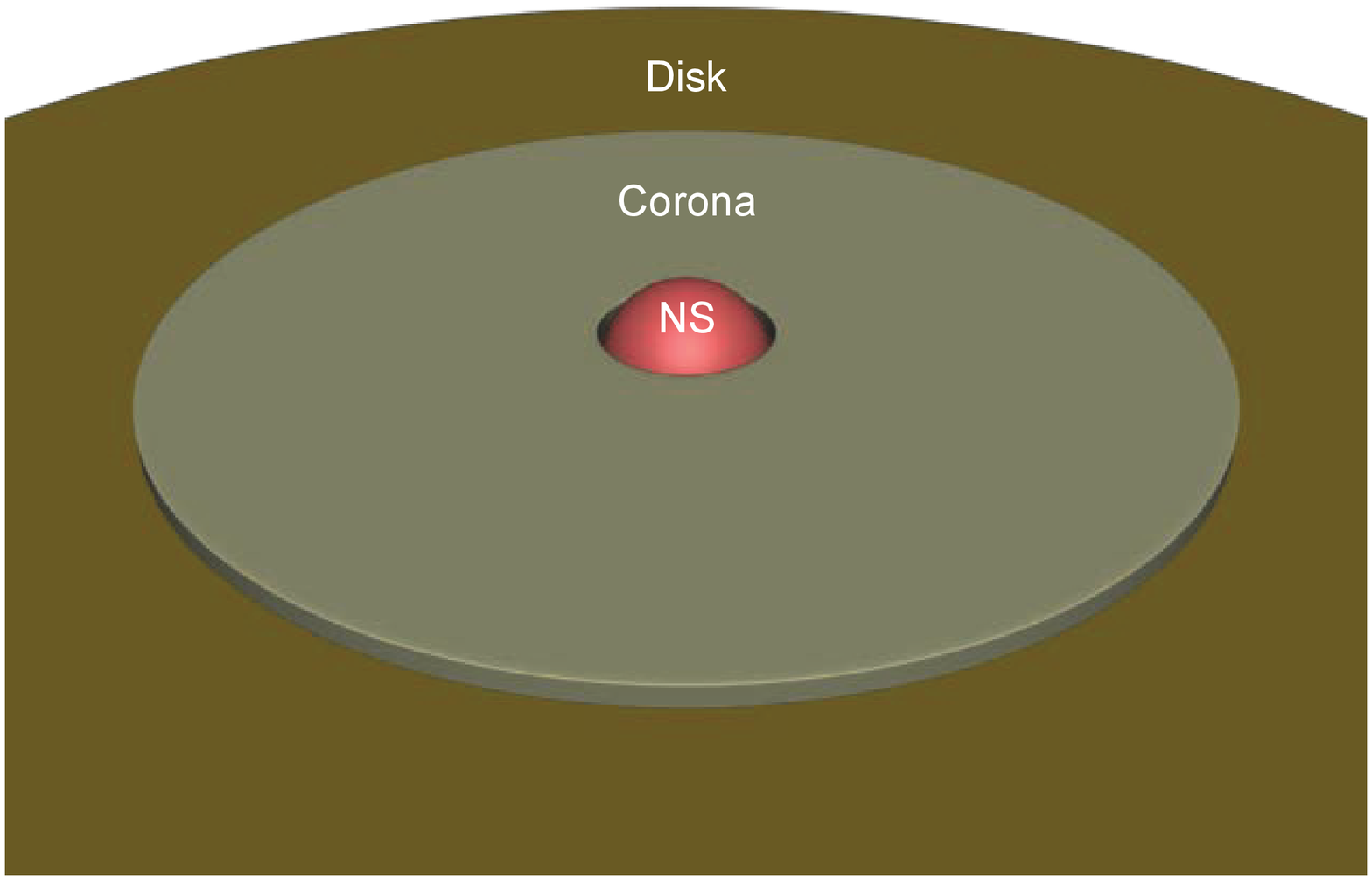}
            \includegraphics[angle=0, scale=0.22]{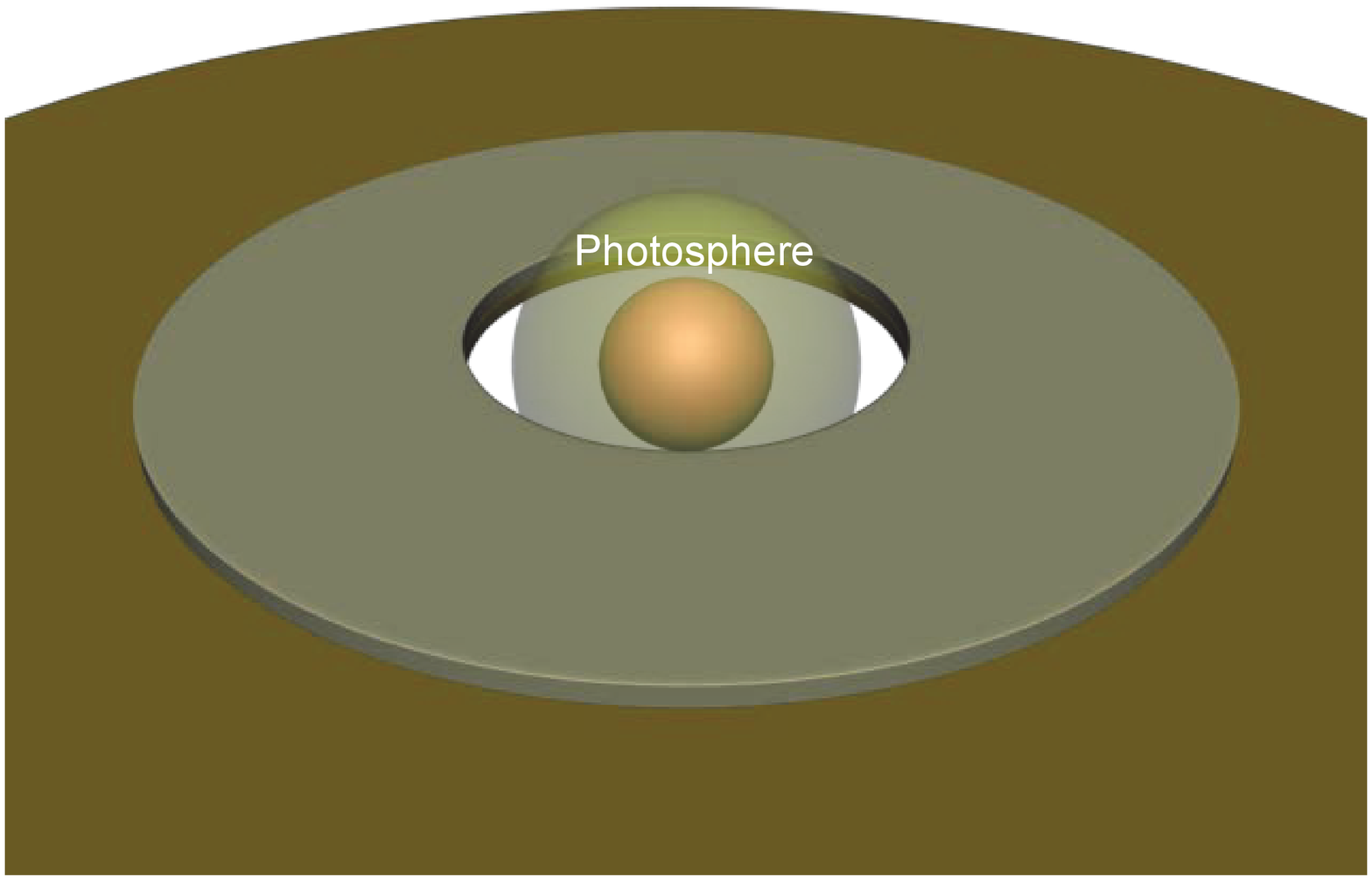}
      \includegraphics[angle=0, scale=0.22]{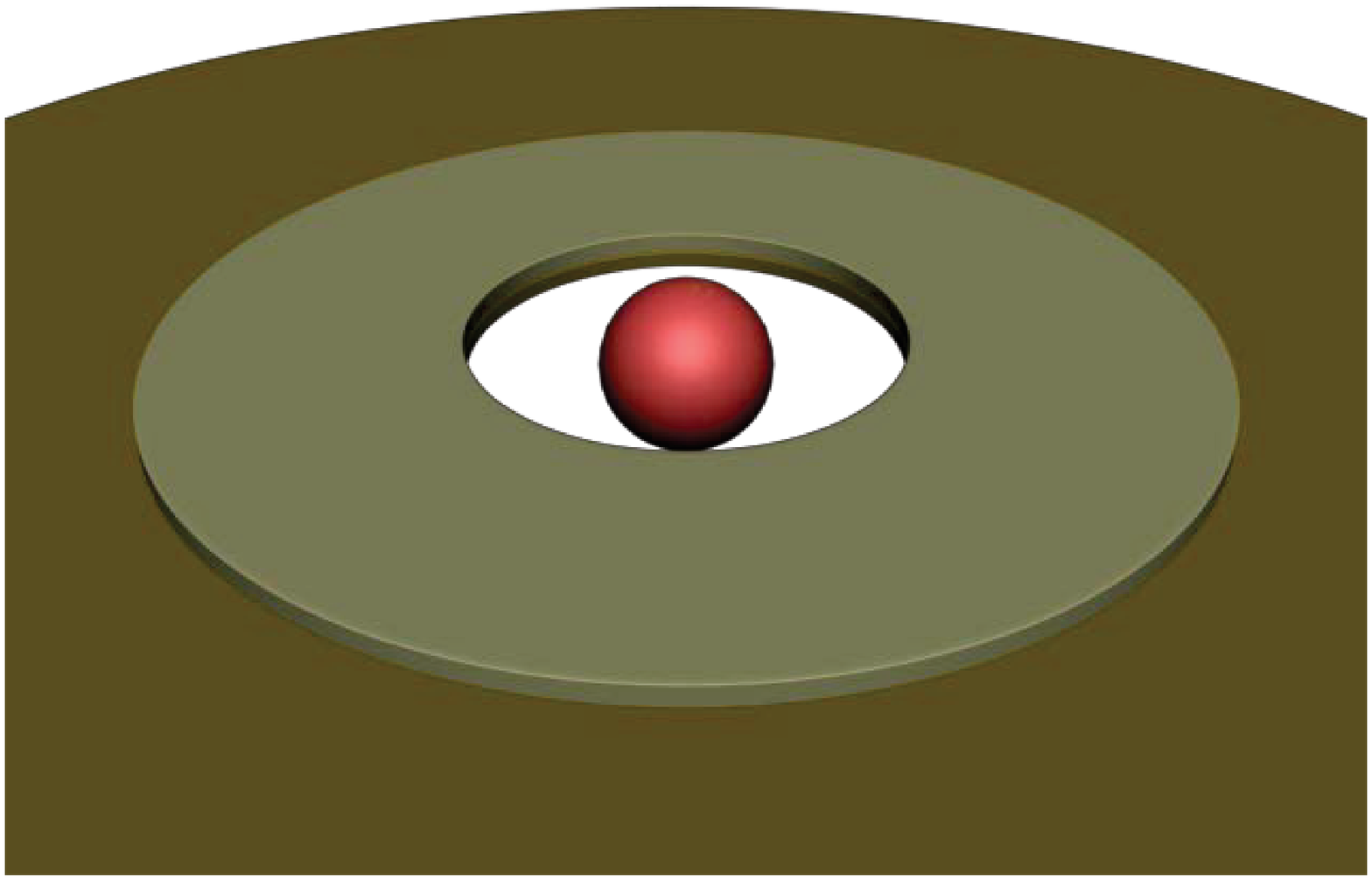}
 \caption{Illustration of the central region of an NS XRB before the PRE phase (left), in the PRE phase (middle) and after the  PRE phase (right) during a burst, in which the inner part of the disk is swept away by the burst in the PRE phase. % radiation pressure.
 }
\label{fig_illustraction}
\end{figure}

The flux-temperature diagram of the burst also indicates that the inner disk radius change causes a bigger  visible part of the NS surface.  If the whole NS surface shows up as a single-temperature blackbody and a constant color correction factor, the burst flux $F$ should scale as $kT_{\rm bb}^{4}$ in the flux-temperature diagram, and the slope represents the emitting area in the double logarithmic coordinates \citep{Guver2012}. As shown in Figure \ref{fig_t_f}, the rising phase and decaying phase obey  different $F \propto kT_{\rm bb}^{4}$. We fit the two sets with $F=\frac{R^2}{D^2}\sigma T^4$, the blackbody radius of the rising and decaying phase is 4.3$\pm$0.11 km and 5.3$\pm$0.05 km with $D=4$ kpc. %Similar results were also reported in PRE   burst of 4U~1728--34 \citep{Shaposhnikov2003} and 4U~1820--30\citep{Shaposhnikov2004}.

Assuming the NS radiates at the Eddington limit in the PRE phase and the disk reaches the NS surface before the PRE phase, the blackbody flux ratio  detected at the rising phase $F_{\rm rise}$ and  decaying phase  $F_{\rm decay}$ is  positively associated with inclination angel $i$, i.e.,$\frac{F_{\rm rise}}{F_{\rm decay}}=(1+ {\rm cos}~i)/2$  \citep{Shaposhnikov2003,Shaposhnikov2004}.
The inclination angel $i$ is estimated as $\sim$ 70$^\circ$.
However, this result is bigger than the value $\sim$ 30$^\circ$--40$^\circ$ derived from the spectral fit results on an outburst of 4U~1608--52 by  a reflection model \citep{Degenaar2015}.
%Since it has a certain probability of a partial occultation of the lower NS hemisphere by the disk at the rising phase, the $F_{\rm rise}$ is a lower limit, which causes an overestimation of the inclination angel. In this case, 70$^\circ$ is an upper limit of the inclination angel.
%\citep{Mondal2017}.

\subsection{Corona/boundary-layer reacting on burst}

The interaction between the burst and inhabited persistent emission was first studied  from the long-duration, brightest and most vigorous PRE bursts with moderate/super expansions in 4U~1820--30 and 4U~1636--536 \citep{Ball2004,KeeK2014b} (a factor of $\sim$10--10$^{4}$ increases in emission area).
Then in short-duration bursts,
%Meanwhile,
this interaction was mainly observed as the persistent spectral change, rather than the burst spectral change, i.e., enhancement accretion rate, deficit at hard X-ray band, reflection by the disk and driven outflow.

Type-I X-ray burst happens on the NS surface, which is also in the accretion environment.
In principle, the burst spectrum may be influenced  due to the Comptonization of the burst photons by the surrounding corona/boundary-layer \citep{chen2019}.
A Comptt component was  reported  in bursts of 4U~1608--52 from RXTE observations above 3 keV \citep{Kajava2017}. However, their approach resembled the $f_{a}$ model since the Comptt  component is added in the spectra fitting.
In this work, the persistent emission is well fitted by a convolution thermal-Comptonization model.
 As a result, given the similarity, for the burst, adapting the convolution  model  with the parameters in the persistent emission fitting but with a blackbody emission, this could also fit the short-duration burst spectra in the PRE phase.
The goodness of the fit are comparable  with the $f_{a}$ model, but with a colder $kT$ and larger $R$ in the middle of the PRE phase.
If this is the case, the radius of the photosphere is underestimated with the canonical blackbody model or the  $f_{a}$ model.

%A similar speculation is that the enhanced  persistent emissions could come from a region with a  radius of $\sim$200 km.
In principle, the interaction between burst and accretion environment, might be expected to have spectral evolution for both of the burst emission and accretion emission during burst, with spectral shape deviation from pure blackbody and  the model of the pre-burst emission.
The short duration and rapid spectral change limits the accumulated time and photon counts, which in turn requires   larger  detection area and broad band energy coverage which may be satisfied by the next generation of Chinese mission of so-called eXTP (enhanced X-ray Timing and Polarimetry mission) \citep{Zhang2019} or  a contemporary joint observation of the burst by NICER and Insight-HXMT.

%is greatly  in our understanding of the burst influence upon the accretion environment.

%\subsection{Conclusions and Outlook}

\acknowledgements
We thank the reviewer for the constructive feedback and comments that greatly improved the quality of this paper.
This work made use of the data and software from the Insight-HXMT
mission, a project funded by China National Space Administration
(CNSA) and the Chinese Academy of Sciences (CAS).
This research has  made use of data and software provided by of data obtained from the High Energy Astrophysics Science
Archive Research Center (HEASARC), provided by NASA’s
Goddard Space Flight Center.
This work is supported by the National Key R\&D Program of China (2021YFA0718500) and the National Natural Science Foundation of China under grants 11733009, U1838201, U1838202, U1938101, U2038101.

\bibliographystyle{plainnat}

\begin{thebibliography}{99}
%\bibitem[Agrawal\& Hasanet  (2016)]{Agrawal2016}Agrawal, Vivek Kumar;\& Hasan, Mohammad, 2016, arXiv:1611.09004
\bibitem[Armas Padilla et al. (2017)]{ArmasPadilla2017}Armas Padilla, M., Ueda, Y., Hori, T., Shidatsu, M., Munoz-Darias, T., 2017, MNRAS, 467, 290
\bibitem[Ballantyne \& Strohmayer (2004)]{Ball2004}Ballantyne, D. R., \& Strohmayer, T. E. 2004, ApJL, 602, L105
%\bibitem[Ballantyne et al. (2004)]{Ballantyne2004}Ballantyne, D. R., \& Strohmayer, T. E. 2004, ApJL, 602, L105
\bibitem[Bhattacharyya et al. (2018)]{Bhattacharyya2018}Bhattacharyya, S. Yadav, J. S.,  Sridhar, Navin, et al. 2018, ApJ, 860, 88
\bibitem[Belian et al. (1976)]{Belian}Belian, R. D., Conner, J. P., \& Evans, W. D. 1976, ApJ, 206, L135
%\bibitem[Chen et al. (2013)]{chen2013} Chen Y. P., Zhang S., Zhang S. N., et al. 2013, ApJL, 777, 9
\bibitem[Chen et al. (2012)]{chen2012} Chen, Y. P., Zhang, S., Zhang, S. N., et al. 2012, ApJL, 752, 34
\bibitem[Chen et al. (2018)]{chen2018} Chen, Y. P., Zhang, S., Qu, J. L., Zhang, S. N., et al. 2018, ApJL, 864, 30
\bibitem[Chen et al. (2019)]{chen2019} Chen, Y. P., Zhang, S., Zhang, S. N., et al. 2019, Journal of High Energy Astrophysics, 24, 23
\bibitem[Cumming(2004)]{Cumming}Cumming, A. \ 2004, Nucl. Phys. B Proc. Suppl., 132, 435
%\bibitem[Czerny et al. (1987)]{Czerny1987}Czerny, M., Czerny, B., \& Grindlay, J. E. 1987, ApJ, 312, 122
%\bibitem[Degenaar et al.(2015)]{Degenaar2015}Degenaar, N., Miller, J. M., Chakrabarty, D., et al., 2015, MNRAS, 451, L85
\bibitem[Degenaar et al.(2015)]{Degenaar2015}Degenaar, N., Miller, J. M., Chakrabarty, D., Harrison, F. A., Kara, E., \& Fabian, A. C. 2015, MNRAS, 451, L85
\bibitem[Degenaar et al.(2018)]{Degenaar2018}Degenaar, N., Ballantyne, D. R., Belloni, T., et al. 2018, SSRv, 214, 15
%\bibitem[Galloway et al. (2006)]{Galloway2006} Galloway D. K., Psaltis D., Muno M. P., Chakrabarty D., 2006, ApJ, 639, 1033
\bibitem[Done et al. (2007)]{Done2007}Done, C., Gierliński, M., Kubota, A. 2007, A\&AR, 15, 1
\bibitem[Fragile et al.(2020)]{Fragile2020}Fragile, P. C., Ballantyne, D. R., \& Blankenship, A. 2020, NatAs, 4, 541
\bibitem[Galloway et al.(2008)]{Galloway}Galloway, D. K., Muno, M. P., Hartman, J. M., et al. \ 2008, ApJS, 179, 360
\bibitem[Galloway et al.(2020)]{Galloway2020}Galloway, D. K., In’t Zand, J., Chenevez, J., et al. 2020, ApJS, 249, 32
\bibitem[G$\ddot{u}$ver et al.(2012)]{Guver2012}G$\ddot{u}$ver, T., Psaltis, D., \& Zel, F. 2012, ApJ, 747, 76
\bibitem[Gilfanov et al. (2003)]{Gilfanov2003}Gilfanov, M.; Revnivtsev, M.; Molkov, S.Gilfanov et al. 2003, A\&A, 410, 217
\bibitem[Grebenev \& Sunyaev (2002)]{Grebenev2002}Grebenev, S. A., \& Sunyaev, R. A. 2002, AstL, 28, 150
%\bibitem[Garc\'{\i}a et al.(2014)]{Garcia}Garc\'{\i}a, J., Dauser, T., Lohfink, A., et al. 2014, ApJ, 782, 76
\bibitem[in't Zand et al. (1999)]{int1999}in't Zand, J. J. M., Heise, J., Kuulkers, E., et al. 1999, A\&A, 347, 891
\bibitem[in't Zand et al. (2013)]{int2013}in't Zand, J. J. M., Galloway, D. K., Marshall, H. L., et al. 2013, A\&A,553, A83
\bibitem[Ji et al. (2013)]{ji2013}Ji, L., Zhang, S., Chen, Y. P., et al. 2013, MNRAS, 432, 2773
%\bibitem[Ji et al. (2014)]{ji2014}Ji, L., Zhang, S., Chen, Y. P., et al. 2014, ApJL, 791, L39
%\bibitem[Ji et al. (2014a)]{ji2014a} Ji L., Zhang S., Chen Y., et al., 2014a, ApJL, 782, 40
%\bibitem[Ji et al. (2014b)]{ji2014b}Ji L., Zhang S., Chen Y., et al., 2014b, ApJ, 791, L39
%\bibitem[Ji et al. (2014c)]{ji2014c}Ji L., Zhang S., Chen Y. P., et al., 2014c, A\&A, 564, A20
%\bibitem[Kajava et al. (2017)]{Kajava2017}Kajava, J. J. E., Koljonen, K. I. I., N$\ddot{a}$ttil$\ddot{a}$, J., Suleimanov, V., \& Poutanen, J. 2017, MNRAS, 472, 78
%\bibitem[Keek et al. (2012)]{Keek2012}Keek, L., Heger, A., \& in't Zand, J. J. M. 2012, ApJ, 752, 150
%\bibitem[Keek et al. (2014a)]{Keek2014a}Keek, L., Ballantyne, D, R., Kuulkers, E. et al. 2014a, ApJ, 789, 121
\bibitem[Kaastra \& Bleeker (2016)]{Kaastra2016}Kaastra, J. S.; Bleeker, J. A. M. 2016, A\&A, 587, A151
\bibitem[Kajava et al. (2017)]{Kajava2017}Kajava, J. J. E., Koljonen, K. I. I., N$\ddot{a}$ttil$\ddot{a}$, J., Suleimanov, V., \& Poutanen, J. 2017, MNRAS, 472, 78

\bibitem[Keek et al. (2014)]{KeeK2014b}Keek, L., Ballantyne, D. R., Kuulkers, E., \& Strohmayer, T. E. 2014b, ApJL, 797, L23
\bibitem[Keek et al. (2018)]{Keek2018}Keek, L., Arzoumanian, Z., Bult, P., et al. 2018, ApJL, 855,4
\bibitem[Keek et al. (2018a)]{Keek2018a}Keek, L., Arzoumanian, Z., Chakrabarty, D., et al. 2018, ApJL, 856,37
%\bibitem[Lamb \& Miller (1995)]{Lamb}Lamb, F. K., \& Miller, M. C. 1995, ApJ, 439, 828
\bibitem[Lewin et al.(1993)]{Lewin}Lewin, W. H. G., van Paradijs, J., \& Taam, R. E. \ 1993, Space Sci. Rev., 62, 223
%\bibitem[Maccarone \& Coppi(2003)]{maccarone2003}Maccarone, T. J. \& Coppi, P. S. \ 2003, A\&A, 399, 1151
%\bibitem[Miller \& Lamb (1993)]{Miller}Miller, M. C., \& Lamb, F. K. 1993, ApJL, 413, L43
%\bibitem[Muno et al. (2000)]{Muno2000}Muno, M. P., Fox, D. W., Morgan, E. H., \& Bildsten, L. 2000, ApJ, 542, 1016
\bibitem[Li et al.(2020)]{Li2020}Li, X. B., Li, X. F., Tan, Y. et al. 2020, JHEA, 27, 64
%\bibitem[Mondal et al. (2017)]{Mondal2017}Mondal, A. S., Pahari, M., Dewangan, G. C., Misra, R., \& Raychaudhuri, B. 2017, MNRAS, 466,  4991
\bibitem[Muno et al. (2001)]{Muno}Muno, M. P., Chakrabarty, D., Galloway, D. K., \& Savov, P. 2001, ApJ, 553, L157
%\bibitem[Penninx et al. (1989)]{Penninx1989}Penninx W., Damen E., Tan J., Lewin W. H. G., van Paradijs J., 1989, A\&A, 208, 146

\bibitem[PenninxW et al. (1989)]{PenninxW1989}PenninxW., Damen E., van Paradijs J., Tan J., Lewin W. H. G., 1989, A\&A, 208, 146

\bibitem[Poutanen et al. (2014)]{Poutanen}Poutanen, J., N$\ddot{ai}$ttil$\ddot{a}$, J., Kajava, J. J. E. et al. 2014, MNRAS, 442, 3777
\bibitem[S{\'a}nchez-Fern{\'a}ndez et al. (2020)]{Sanchez2020}S{\'a}nchez-Fern{\'a}ndez C., Kajava J. J. E., Poutanen J., et al., 2020, A\&A, 634, A58
\bibitem[Remillard et al. (2022)]{Remillard2022}Remillard, R. A., Loewenstein, M., Steiner, J. F. et al. 2022, AJ, 163, 130
\bibitem[Suleimanov \& Poutanen (2006)]{Suleimanov2006}Suleimanov, V., \& Poutanen, J. 2006, MNRAS, 369, 2036
\bibitem[Shaposhnikov et al. (2003)]{Shaposhnikov2003}Shaposhnikov, N., Titarchuk, L., Haberl, F., 2003, ApJL, 593, L35
\bibitem[Shaposhnikov \& Titarchuk (2004)]{Shaposhnikov2004}Shaposhnikov, N., \& Titarchuk, L. 2004, ApJ, 606, L57
\bibitem[Stahl et al. (2013)]{Stahl}Stahl, A., Klu$\acute{z}$niak, W., Wielgus, M., \& Abramowicz, M. 2013, A\&A, 555, A114
\bibitem[Strohmayer \& Bildsten(2006)]{Strohmayer} Strohmayer, T., \& Bildsten, L.   \ 2006, New views of thermonuclear bursts (Compact stellar X-ray sources), 113, 156
%\bibitem[Suleimanov et al. (2011)]{Suleimanov2011}Suleimanov, V.; Poutanen, J.; Werner, K. 2011, A\&A, 527, A139
%\bibitem[Swank et al. (1976)]{Swank1976}Swank J. H., Becker R. H., Pravdo S. H., Saba J. R., Serlemitsos P. J., 1976a, IAU Circ., 3000, 5
%\bibitem[van Paradijs et al. (1990)]{van1990} van Paradijs J., van der Klis M., van Amerongen S., et al. 1990, A\&A, 234, 181
%\bibitem[Walker  (19922)]{Walker}Walker, M. A. 1992, ApJ, 385, 642
%\bibitem[Watts  (2012)]{Watts}Watts, A. L. 2012, ARA\&A, 50, 609
\bibitem[Suleimanov et al. (2011)]{Suleimanov2011}Suleimanov, V., Poutanen, J., Werner, K. 2011, A\&A, 527, A139
\bibitem[Suleimanov et al. (2012)]{Suleimanov2012}Suleimanov, V., Poutanen, J., Werner, K. 2012, A\&A, 545, A120
\bibitem[Suleimanov et al. (2018)]{Suleimanov2018}Suleimanov, V., Poutanen, J., Werner, K. 2018, A\&A, 619, A114

\bibitem[Wilms et al. (2000)]{Wilms2000}Wilms, J., Allen, A., \& McCray, R. 2000, ApJ, 542, 914
\bibitem[Worpel et al. (2013)]{Worpel2013}Worpel, H., Galloway, D. K., \& Price, D. J. 2013, ApJ, 772, 94
\bibitem[Worpel et al. (2015)]{Worpel2015}Worpel, H., Galloway, D. K., \& Price, D. J. 2015, ApJ, 801, 60
%\bibitem[Zhang et al. (2011)]{Zhang2011} Zhang G. B., M\'endez M., Altamirano D., 2011, MNRAS, 413, 1913Z
\bibitem[Zhang et al. (2000)]{Zhang2000}Zhang, S. N., et al., 2000, Science, 287, 1239
\bibitem[Zhang et al. (2014)]{Zhang2014} Zhang, S.,  L,u F. J., Zhang, S. N. et al. in Space Telescopes and Instrumentation 2014: Ultraviolet to Gamma Ray, Proc. SPIE, Vol. 9144 (2014) p. 914421
\bibitem[Zhang et al. (2019)]{Zhang2019} Zhang, S. N., Santangelo, A., Feroci, M., et al.  2019, Science China Physics, Mechanics \& Astronomy, Volume 62, Issue 2, article id. 29502, 25
\bibitem[Zhang et al. (2020)]{Zhang2020}Zhang, S.-N., Li, T.-P., Lu, F.-J., et al. 2020, SCPMA, 63, 249502
%\bibitem[$\dot{Z}$ycki et al. (1999)]{zycki1999} $\dot{Z}$ycki, P. T., Done,  Chris, Smith, D. A. 1999, MNRAS, 309, 561
\bibitem[Zdziarski et al. (2021)]{Zdziarski2021}Zdziarski, A. A., Szanecki, M., Poutanen, J., Gierlinski, M., \& Biernacki, P. 2021, MNRAS, 492, 5234

%     in't Zand, J. J. M., Galloway, D. K., Marshall, H. L., et al. 2013, A\&A,553, A83
%    Cumming, A. \ 2004, Nucl. Phys. B Proc. Suppl., 132, 435
 %   Galloway, D. K., Muno, M. P., Hartman, J. M., et al. \ 2008, ApJS, 179, 360
%    Lewin, W. H. G., van Paradijs, J., \& Taam, R. E. \ 1993, Space Sci. Rev., 62, 223

\end{thebibliography}

\begin{landscape}

\begin{table}[ptbptbptb]
\begin{center}
\caption{The bursts obsid and peak time of 4U~1608--52  detected by Insight/HXMT in 2020 ourbutst }
\label{tb}
 %\vspace{5pt}
\begin{tabular}{cccccccccccccccccc}
\\\hline
   obsid  &Start Time & Elapsed Time (s) &	LE GTI (s) & ME GTI (s) 	& Burst peak time (MJD) \\\hline
 	P030402100401$^{\mathrm{*}}$	& 59069.63746  (2020-08-08T15:16:50) & 14708 &	2480 &3750 &59069.77077	\\\hline
  	P030402100602	& 59079.83751  (2020-08-18T20:04:55) & 11617 &	120 &480 &59079.85655	\\	
\hline
\end{tabular}
\end{center}
\begin{list}{}{}
\item[${\mathrm{*}}$]{This work.}
\item[Note:]{\textbf{Both  bursts are detected by the three payloads of Insight-HXMT. However, the second bursts, which is only half bright as the first one, did not located the good-time-interval.} }
\end{list}
\end{table}

\begin{table}[ptbptbptb]
%\tiny
%\scriptsize
%\footnotesize
\begin{center}
\caption{The NICER obsid at the same day when Insight-HXMT detected the burst }
 %\vspace{5pt}
 \label{table_hxmt}
\begin{tabular}{cccccccccccc}
\\\hline
obsid &  Start Time & Elapsed Time (s) &	GTI (s)    \\\hline
3657026501  & 59069.61551 (2020-08-08T14:43:40)& 3600   &1707\\\hline
\hline
\end{tabular}
\end{center}
%\begin{list}{}{}
%\item[${\mathrm{a}}$]{: The bolometric flux of the blackbody $F_{\rm bb}$ is in unit of $10^{-8}~{\rm erg/cm}^{2}/{\rm s}$.}
%\item[${\mathrm{b}}$]{fitting reduced $\chi^{2}$}
%\item[${\mathrm{***}}$]{burst oscillation detected.}
%\item[Note:]{The OBSID, time,  peak flux  in 2-10 keV (in units of cts/s), PRE or non-PRE bursts  are given.}
%\end{list}
\end{table}

\begin{table}
\centering
\caption{The results of the spectral fit of the LE, ME and NICER spectra in the 0.4--20 keV range   with  cons*tbabs*thcomp*diskbb}
\label{persist_fit}
\vskip -0.4cm
\begin{tabular}{ccccccc}
\\\hline
$N_{\rm H}$  & $\tau$ & $kT_{\rm e}$  & $f_{\rm sc}$ & $kT_{\rm in}$
 & $N_{\rm diskbb}$& $\chi_\nu^2$ \\
$10^{22}~{\rm cm}^{-2}$ &  & keV & &keV & $10^2$ &   \\
\hline
$1.33_{-0.01}^{+0.01}$ & $10.2^{+0.3}_{-0.4}$ & $3.02^{+0.08}_{-0.08}$ & $0.76_{-0.05}^{+0.05}$  & $0.68_{-0.02}^{+0.01}$ & $8.38_{-0.96}^{+1.03}$ & 802/846 \\ %0.4-20 keV
%%%$1.05_{-0.01}^{+0.01}$ & $10.3^{+0.2}_{-0.3}$ & $3.03^{+0.08}_{-0.07}$ & $0.75_{-0.04}^{+0.04}$  & $0.64_{-0.01}^{+0.01}$ & $12.1_{-1.2}^{+1.7}$ & 1399/1441 \\ %1-20 keV

\hline
\end{tabular}
\end{table}

\end{landscape}

\clearpage

% \begin{figure}[t]
 %  \includegraphics[angle=270, scale=0.5]{lc_nicer_le_me.eps}
% \caption{
% Lightcurves of   4U~1608--52 with time bin 1 s. The top, middle and bottom panel is NICER, LE and ME   in their full energy band respectively. No background is subtracted.  }
%\label{fig_lc_nicer_le_me}
%\end{figure}

\clearpage

\clearpage

\end{document}